\pdfoutput=1
\documentclass[11pt]{article}

\usepackage{emnlp2021}

\usepackage{latexsym}

\usepackage[T1]{fontenc}
\usepackage[utf8]{inputenc}

\usepackage{amsmath,amsfonts,bm}

\def\figref#1{Figure~\ref{#1}}
\def\eqref#1{equation~\ref{#1}}
\def\1{\bm{1}}

\def\vh{{\bm{h}}}

\def\vn{{\bm{n}}}

\def\vx{{\bm{x}}}
\def\vy{{\bm{y}}}

\DeclareMathAlphabet{\mathsfit}{\encodingdefault}{\sfdefault}{m}{sl}
\SetMathAlphabet{\mathsfit}{bold}{\encodingdefault}{\sfdefault}{bx}{n}

\usepackage{times}
\usepackage{url}
\usepackage{etoolbox,multirow,adjustbox}
\usepackage{wrapfig,lipsum,booktabs}
\usepackage{hyperref}       % hyperlinks
\usepackage{url}            % simple URL typesetting
\usepackage{booktabs}       % professional-quality tables
\usepackage{nicefrac}       % compact symbols for 1/2, etc.
\usepackage{microtype}      % microtypography
\usepackage{multirow}
\usepackage{subfig}
\usepackage{enumerate}
\usepackage{caption}
\usepackage{graphicx}
\usepackage{xspace}
\usepackage{paralist}
\usepackage{float}
\usepackage{scrextend}
\usepackage[tight-spacing=true]{siunitx}
\usepackage{color,soul}
\usepackage{etoolbox,multirow,adjustbox}
\usepackage{comment}
\usepackage{enumitem,kantlipsum}
\usepackage{xcolor}
\usepackage{makecell}

\newlength{\oldintextsep}
\usepackage{amsthm}
\usepackage{algpseudocode,algorithm}
\usepackage{tabularx,adjustbox}
\newcolumntype{L}[1]{>{\hsize=#1\hsize\raggedright\arraybackslash}X}%
\newcolumntype{R}[1]{>{\hsize=#1\hsize\raggedleft\arraybackslash}X}%
\newcolumntype{C}[1]{>{\hsize=#1\hsize\centering\arraybackslash}X}%
\usepackage{todonotes}
\usepackage{breqn}
\clearpage
\extrafloats{100}
\RequirePackage{tcolorbox}
\tcbuselibrary{breakable}
\tcbset{parbox=false, left=1ex, right=1ex}

\newcommand{\ie}{\textit{i.e.},\xspace}
\newcommand{\eg}{\textit{e.g.},\xspace}

\def\eg{{\em e.g.,}\xspace}
\def\ie{{\em i.e.,}\xspace}
\def\cf{{\em c.f.,}\xspace}

\newcommand{\tabref}[1]{Table~\ref{#1}\xspace}

\newcommand{\resource}[1]{\textsc{#1}}

\title{Killing One Bird with Two Stones: Model Extraction and Attribute Inference Attacks against BERT-based APIs}
\author{Chen Chen \textsuperscript{\rm 1}\thanks{\quad Equal contribution.} , Xuanli He\textsuperscript{\rm 2}$^*$, Lingjuan Lyu\textsuperscript{\rm 3}\thanks{\quad Corresponding author.} , Fangzhao Wu\textsuperscript{\rm 4}
\\
\textsuperscript{\rm 1} Zhejiang University
\textsuperscript{\rm 2} Monash University
\textsuperscript{\rm 3} Sony AI
\textsuperscript{\rm 4} Microsoft Research Asia
}

\begin{document}
\maketitle
%%%%%%%%% ABSTRACT
\begin{abstract}
The collection and availability of big data, combined with advances in pre-trained models (e.g., BERT, XLNET, etc), have revolutionized the predictive performance of modern natural language processing tasks, ranging from text classification to text generation.
This allows corporations to provide machine learning as a service (MLaaS) by encapsulating fine-tuned BERT-based models as APIs. However, BERT-based APIs have exhibited a series of security and privacy vulnerabilities. For example, prior work has exploited the security issues of the BERT-based APIs through the adversarial examples crafted by the extracted model.
However, the privacy leakage problems of the BERT-based APIs through the extracted model have not been well studied.
On the other hand, due to the high capacity of BERT-based APIs, the fine-tuned model is easy to be overlearned, but what kind of information can be leaked from the extracted model remains unknown.
In this work, we bridge this gap by first presenting an effective model extraction attack, where the adversary can practically steal a BERT-based API (the target/victim model) by only querying a limited number of queries. We further develop an effective attribute inference attack which can infer the sensitive attribute of the training data used by the BERT-based APIs. Our extensive experiments on benchmark datasets under various realistic settings validate the potential vulnerabilities of BERT-based APIs. Moreover, we demonstrate that two promising defense methods become ineffective against our attacks, which calls for more effective defense methods. 
\end{abstract}

\begin{sloppypar}
%%%%%%%%% BODY TEXT
\section{Introduction}
The emergence of \textbf{B}idirectional \textbf{E}ncoder \textbf{R}epresentations from \textbf{T}ransformers (BERT)~\cite{devlin2018bert} has revolutionized the natural language processing (NLP) area, leading to state-of-the-art performance on a wide range of NLP tasks with minimal task-specific supervision. At the same time, with the increasing success of contextualized pre-trained representations for transfer learning, powerful NLP models can be easily built by fine-tuning the pre-trained models like BERT or XLNet~\cite{yang2019xlnet}. Building NLP models on the pre-trained representations typically only require several task-specific layers on top of BERT~\cite{nogueira2019passage,liu2019text,
joshi2020spanbert}. Commercial companies often offer machine learning models as a service on their cloud platforms. Examples include Google Prediction API\footnote{https://cloud.google.com/prediction}, Microsoft Azure Machine Learning (Azure ML)\footnote{https://studio.azureml.net},  Amazon Machine Learning (Amazon ML)\footnote{https://aws.amazon.com/machine-learning}.
In terms of NLP domains, NLP models such as task-specific BERT models are often made indirectly accessible through pay-per-query prediction APIs~\cite{krishna2019thieves}, \ie the only information the attacker can access is the model prediction.  

However, prior works have shown that existing NLP APIs are vulnerable to model extraction or stealing attack, which can reconstruct a copy of the remote NLP model based on the carefully-designed queries and the outputs of the target API \cite{krishna2019thieves,wallace2020imitation}, violating the intellectual property of the target API. Pretrained BERT models further make it easier to apply model extraction attacks on the specialized NLP models obtained by fine-tuning the pretrained BERT models \cite{krishna2019thieves}. This is because using a pretrained BERT gives attackers a significant head-start endowed with the knowledge and properties of language~\cite{krishna2019thieves}, and only the final layer of the model needs to be randomly initialized. In fact, model extraction/stealing is not just hypothetical, it is a practical concern for industrial NLP systems. The enormous commercial benefit allures competing companies or individual users to extract or steal these successful APIs. For instance, competing companies in NLP business have been caught stealing models\footnote{https://googleblog.blogspot.com/2011/02/microsofts-bing-uses-google-search.html}. Even worse, the attacker could potentially surpass victims by conducting unsupervised domain adaptation and multi-victim ensemble~\cite{xu2021beyond}. Although API providers can conduct a watermarking method for intellectual property (IP) protection~\cite{he2021protecting}, it still belongs to post-hoc operations, which cannot prevent model extraction at its first stage.

Besides model extraction that may cause the violation of the intellectual property, it is of paramount importance to investigate what kind of information can be gained or exploited from the extracted model. Some recent works have showed that 
the adversarial examples crafted by the extracted model could be transferred to the black-box victim model~\cite{wallace2020imitation,he2021model}.
\textbf{However, information leakage from the extracted model has received little attention in the literature}. 

Our work fills in this gap and investigates the vulnerability of BERT-based APIs through a two-stage attack. We first launch a model extraction attack, where the adversary queries the target model with the goal to steal it and turn it into a white-box model. With the extracted model, we further demonstrate that it is possible to infer the sensitive attribute of the training data used by the target API. All these attacks will seriously infringe the interests of the original API owner. To the best of our knowledge, this is the first attempt that investigates privacy leakage from the extracted model. Our results highlight the risks of using the pre-trained BERT to deploy the task-specific APIs through the lens of model extraction attack and attribute inference attack\footnote{Code will be available upon publication}. Moreover, we show that our attacks can mostly evade the investigated defense strategies, unless when the defense strategies compromise the main task utility significantly. We remark that although our work leverages BERT-based models as a case study, both the methodology and discovered issues are universal and model-agnostic.

\section{Related Work}
Model extraction attack (also referred to as ``stealing" or ``reverse-engineering") aims to steal an intellectual model from cloud services~\cite{tramer2016stealing, orekondy2019knockoff, krishna2019thieves, wallace2020imitation}.
Model extraction attack has been studied both empirically and theoretically, for simple classification tasks~\cite{tramer2016stealing}, vision tasks~\cite{orekondy2019knockoff},
and NLP tasks~\cite{krishna2019thieves, wallace2020imitation}. As opposed to stealing parameters~\cite{tramer2016stealing}, hyper-parameters~\cite{wang2018stealing}, architectures~\cite{oh2019towards}, and decision boundaries~\cite{tramer2016stealing,papernot2017practical}, in this work, we attempt to create a local copy or steal the functionality of a black-box victim model~\cite{krishna2019thieves,orekondy2019knockoff}, i.e., a model that replicates the performance of the victim model as closely as possible. If reconstruction is successful, the attacker has effectively stolen the intellectual property. 

Furthermore, the extracted model could be used as a reconnaissance step to facilitate later attacks~\cite{krishna2019thieves}. For instance, the adversary could construct adversarial examples that will force the victim model to make incorrect predictions~\cite{wallace2020imitation,he2021model}. However, to the best of our knowledge, none of the previous works investigate whether the extracted model can facilitate private information inference about the training data of the victim model.

\begin{figure*}
    \centering
    \includegraphics[width=6.5in]{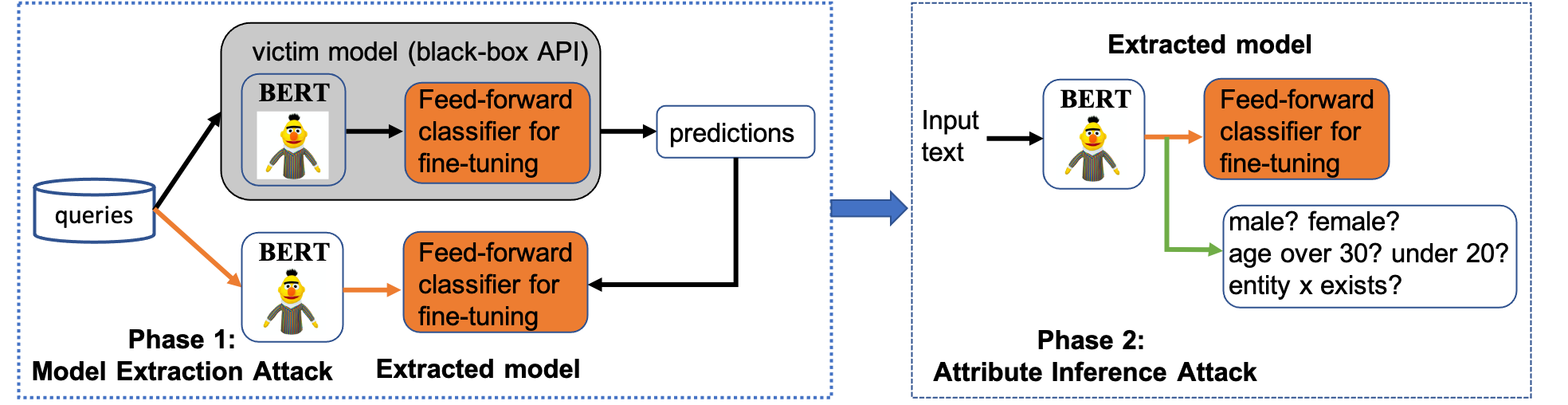}
    \caption{The workflow of our proposed attacks. Phase 1: model extraction attack. We first sample queries, label them using the victim API, then train an extracted model on the resulting data. Phase 2: attribute inference attack based on the extracted model. For attribute inference attack, we train an extra attribute inference model to infer the demographic attributes from BERT representation on any input text.}
    \label{fig:attack_pipeline}
\end{figure*}

\citet{fredrikson2014privacy} first proposed \emph{model inversion attack} in order to infer missing attributes of an input feature vector based on the interaction with a trained ML model. In NLP domain, the input text often provides sufficient clues to portray the author, such as gender, age, and other important attributes. For example, sentiment analysis tasks often have privacy implications for authors whose text is used to train models.
Prior works~\cite{coavoux2018privacy} have shown that user attributes can be easily detectable from online review data, as used extensively in sentiment analysis~\cite{hovy2015user}. One might argue that sensitive information like gender, age, location and password may not be explicitly included in model predictions.
Nonetheless, model predictions are produced from the input text, which can meanwhile encode personal information that might be exploited for adversarial usages, especially considering the fact that a modern deep learning model owns more capacity than the need to perform well on their main tasks~\cite{zhang2017understanding}.
On the other hand, the naive solution of removing protected attributes is insufficient: other features may be highly correlated with, and thus predictive of, the protected attributes~\cite{pedreshi2008discrimination}.

\section{Attacking BERT-based API}
In this work, we consider an adversary attempting to extract or steal BERT-based APIs for financial gain, followed by private information exploitation from the extracted model. As shown in Figure~\ref{fig:attack_pipeline}, the whole attack pipeline against BERT-based APIs can be summarized into two phases. In phase 1, we launch \emph{model extraction attack} (MEA) by first sending the sampled queries to the victim API for labeling, then we train an extracted model on the labeled queries. In phase 2, we conduct \emph{attribute inference attack} (AIA) based on the extracted model. We empirically validate that the extracted model can help \emph{enhance} privacy leakage in Section~\ref{sec:aia}.

We remark that our attack pipeline is applicable to a wide range of BERT-based APIs, as we assume: (a) the capabilities of the attacker are limited to observing only the model outputs from the APIs; (b) the number of queries is limited, avoiding sounding alarms with the API provider and risking an account ban; (c) the query distribution and the training data distribution of the victim  model could be different; (d) the architectures between the extracted model and the victim model could be different.

\subsection{Victim Model: BERT-based API}
Throughout this paper, we mainly focus on the BERT-based API as the victim model, which is generally deployed as a commercially available black-box API. BERT produces rich natural language representations which transfer well to most downstream NLP tasks. Many NLP systems are typically based on a pre-trained BERT~\cite{devlin2018bert, liu2019multi, nogueira2019passage, joshi2020spanbert} by leveraging the fine-tuning methodology, which adds a few task-specific layers on top of the publicly available BERT base,\footnote{\url{https://github.com/google-research/bert}} and fine-tunes the whole model. 

Previous work has shown that owing to the over-learning, the private information can be encoded into BERT embedding~\cite{song2020information}, posing a threat of information leakage. However, \cite{song2020information} directly relied on the embeddings of a target sample obtained from the target model to predict the sample's target attributes. Thus, the adversary is assumed to have \textbf{white-box} access to the target model. We argue that this setting is not applicable to the model stealing setting, in which the attacker only has \textbf{black-box} access to the target model. However, we still believe this undesired privacy leakage can be spread to the extracted model as well, \ie, an attacker can infer the private information related to the victim training data from the hidden representation generated by the extracted BERT.

\subsection{MEA against BERT-based API}
In model extraction attack, an adversary with black-box query access to the victim model attempts to reconstruct a local copy (``extracted model'') of the victim model.
In a nutshell, we perform model extraction attack in a transfer learning setting, where both the adversary and the victim model fine-tune a pretrained BERT. Generally, MEA can be formulated as a two-step approach, as illustrated by the left part in Figure~\ref{fig:attack_pipeline}: 
\begin{enumerate}[wide, labelwidth=!, labelindent=0pt]
\item %Step 1: 
Attacker crafts a set of inputs as queries (transfer set), then sends them to the victim model (BERT-based API) to obtain predictions\footnote{For classification tasks, many APIs including Google Cloud and IBM Cloud can provide prediction scores.} or hard labels;

\item Attacker reconstructs a copy of the victim model as an ``extracted model'' by training on query-prediction pairs. 
\end{enumerate}

Since the attacker does not have the original training data of the victim model, we need to sample $m$ queries $\{\vx_{i}\}_{i=1}^{m}$ either from the same distribution or different distributions from the training data of the victim model. For each $\vx_{i}$, the victim model returns a $K$-dim posterior probability vector $\vy_{i} \in [0,1]^K$, $\sum_k y_i^k=1$. The resulting dataset $\{\vx_{i}, \vy_{i}\}_{i=1}^{m}$ is used to train the extracted model. Once the attacker extracts or steals the victim model successfully, the attacker does not need to pay the provider of the original API anymore for the prediction of new data points. Even worse, the extracted model can be sold at a cheaper price for commercial benefits, eroding the market shares of the victim API. 

\subsection{AIA against BERT-based API}
\label{sec3:AIA}
Next, we investigate how to leverage the extracted model to aid the sensitive attribute inference of the training data used by the victim API, \ie \emph{attribute inference attack} (AIA). For any record $\vx$, we remark that AIA is different from inferring the attribute distribution $p(x^s|y, x^{ns})$ in model inversion attack~\cite{yeom2018privacy}. AIA aims to reconstruct the exact sensitive attribute value $x^s$, which corresponds to the i-th sensitive attribute in the record $\vx$, \ie $x[i]$. Henceforth, we can rewrite it as $\vx=[x^{ns},x^s]$, where $x^{ns}=[x[:i-1],x[i+1:]]$ represents the non-sensitive attributes, while $x^s=x[i]$ represents the sensitive/target attribute. Our intuition behind AIA is that the BERT representation generated by the extracted model can be used to facilitate the inference of the sensitive attribute of the training data of the victim model~\cite{li2018towards,coavoux2018privacy}. Note that in our work, the \textbf{only} explicit information that is accessible to the attacker is model prediction given by the victim model, rather than the original BERT representation. We specifically exploit the BERT representation of the extracted model, as it generally encodes more informative message for the follow-up classification~\cite{song2020information}.

\begin{algorithm}[t]
\caption{Attribute Inference Attack}%Sensitive 
\label{Algorithm:aia}
\begin{algorithmic}[1]
\State \textbf{Input:} extracted model $g'_V$, attribute labeled auxiliary data $D_{aux}$, BERT representation layer $\vh$,
non-sensitive attributes of any record $x_*^{ns}$
\State Query $g'_V$ with $D_{aux}$ and collect $\{(\vh(x_i), s_i)| x_i \in D_{aux}\}$
\State Train an inference model $f$ that predicts $s$ 
on $\{\vh(x_i), s_i\}$.
\State Query $g'_V$ with $x_*^{ns}$ to get the target BERT representation $\vh(x_*^{ns})$
\State\Return $\hat{s}=f(\vh(x_*^{ns}))$
\end{algorithmic}
\end{algorithm}

The main algorithm for AIA is given in Algorithm~\ref{Algorithm:aia}.
Once the extracted model $g'_V$ is built, we feed a limited amount of auxiliary data $D_{aux}$ with labelled attribute into $g'_V$ to collect the BERT representation $\vh(x_i^{ns})$ for each $x_i \in D_{aux}$. For each sensitive attribute of interest, a specific inference model (\cf Section~\ref{sec:aia}) is trained on 
$\{\vh(x_i, s_i)\}$, in order to infer the sensitive attribute; in our case, they are gender, age and named entities (\cf\tabref{tab:data}).
During test time, attacker can first derive the BERT representation of any record by using the extracted model, then feed the extracted BERT representation into the trained inference model to infer the sensitive attribute. In the case when the attacker knows that a particular sample was involved in the training process of the victim model, \ie its membership, the attacker can successfully infer its sensitive attributes given its non-sensitive attributes, causing privacy leakage of the training data of the victim model.

\section{Experiments and Analysis}

\subsection{NLP Tasks and Datasets}
\begin{table}[!htp]
\scalebox{0.7}{
    \centering
% \resizebox{\columnwidth}{!}{
    \begin{tabular}{llllll}
    \hline
Dataset & Private Variable & \#Train & \#Dev& \#Test & Category  \\
    \toprule
%      en-de &4.2M & 3000 & 3003\\
\resource{tp-us} & age, gender & 22,142  & 2,767 & 2,767 & SA \\
\resource{ag} & entity & 11,657 & 1,457 &1,457  & TC\\
\resource{ag} full & - & 112k & 1,457 & 1,457 & TC\\
Blog & age, gender & 7,098 &887 &887 & TC\\
Yelp & - & 520k & 40,000 & 1,000 & SA \\
      
     \bottomrule
 \end{tabular}}
%  }
    % \vspace{3pt}
    \caption{Summary of NLP datasets. SA: sentiment analysis; TC: topic classification.}
    \label{tab:data}
\end{table}

As shown in \tabref{tab:data}, we experiment on four NLP datasets that
focus on two tasks: sentiment analysis and topic classification\footnote{Google Cloud and Microsoft Azure both support sentiment classification and topic classification.}. For all experiments, we report the mean result over five trials.

\textbf{Trustpilot (TP)}. Trustpilot Sentiment dataset~\citep{hovy2015user} contains reviews associated with a sentiment score on a five point scale, and each review is associated with 3 attributes: gender, age and location, which are self-reported by users. The original dataset is comprised of reviews from different locations, however, in this paper, we only derive \resource{tp-us} for study. Following \citet{coavoux2018privacy}, we extract examples containing information of both gender and age, and treat them as the private information.
We categorize ``age'' into two groups: ``under 34'' (\resource{u34}) and ``over 45'' (\resource{o45}).

\textbf{AG news}. \resource{ag} news corpus~\citep{del2005ranking} aims to predict the topic label of the document, with four different topics in total. Following \citep{zhang2015character, jin2019bert}, we use both ``title'' and ``description'' fields as the input document.
We use full \resource{ag} news dataset for MEA, which we call \resource{ag} news (full). As AIA %requires 
takes the entity as the sensitive information, we use the corpus filtered by~\citet{coavoux2018privacy}, which we call \resource{ag} news. The resultant \resource{ag} news merely includes sentences with the five most frequent person entities, and each sentence contains at least one of these named entities. Thus, the attacker can identify these five entities as five independent binary classification tasks.

\textbf{Blog posts (Blog)}. We derive a blog posts dataset from the blog authorship corpus presented in~\citep{schler2006effects}. We recycle the corpus preprocessed by ~\citet{coavoux2018privacy}, which covers 10 different topics. Similar to \resource{TP-US}, the private variables are comprised of the age and gender of the author. In particular, the age attribute is binned into two categories, ``under 20'' (\resource{U20}) and ``over 30'' (\resource{O30}).

\textbf{Yelp Polarity (Yelp)}. Yelp dataset is a document-level sentiment classification \citep{zhang2015character}. The original dataset is in a five point scale (1-5), while the polarized version assigns negative labels to the rating of 1 and 2 and positive ones to 4 and 5.  

\subsection{Data Partition and Research Questions}
\label{sec:data_partition}

For each dataset in \tabref{tab:data}, we first allocate only 10\% data as one type of the auxiliary data $D_{aux}$, \ie same-domain auxiliary data, we leave the discussion of cross-domain auxiliary data in Section~\ref{sec:aia}. For the remaining data, we randomly split the whole dataset into two halves $D=\{D_V,D_{Q}\}$, where $|D_V|=|D_{Q}|$. The first half (denoted as $D_V$) is used to train the victim model, whereas the second half (denoted as $D_{Q}$) is specifically reserved as the public data for querying the APIs for MEA purpose (\cf \tabref{tab:mea_full}). Note that there is no overlap between $D_V$ and $D_{Q}$, and the assumption of the same distribution between the public data and the original training data is a common practice~\cite{song2020information}, and the adversary might be able to capture some public data through external means. Overall, our attack experiments are mainly designed to answer the following research questions (\textbf{RQ}s),

\textbf{RQ1}: Will the attacker be able to extract a model that is comparable with the original victim model by using limited queries from either the same or different distribution?

\textbf{RQ2}: Can we infer the sensitive attribute of the training data of the victim model from the extracted model? Does the extracted model help accelerate privacy leakage?

\begin{table*}[h]

    \centering
    \begin{tabular}{lcc|cc|cc|cc}
    \hline
Query   &   \multicolumn{2}{c}{AG news (full)} & \multicolumn{2}{c}{Blog} & \multicolumn{2}{c}{TP-US} & \multicolumn{2}{c}{Yelp}\\
         \toprule
         
    &     uni-gram & 5-gram &uni-gram & 5-gram & uni-gram & 5-gram &uni-gram & 5-gram \\
           \midrule
reviews   & 68.22\% &0.53\% & 47.21\% & 0.73\% & 60.86\% & 2.57\%\ & 52.68\% & 2.64\% \\
       
news  & 72.13\% & 1.24\% & 44.76\% & 0.06\% & 51.28\% & 0.12\% & 38.69\% & 0.19\% \\
    \hline
    \end{tabular}
    \vspace{3pt}
    \caption{Percentage of uni-gram and 5-gram recall-based overlap between different queries and test sets. Since AG news is derived from AG news (full) and they show a similar distribution, we omit it.%Note the high accuracies for some tasks even at low query budgets, and diminishing accuracy gains at higher budgets.
    }
    %\vspace{-5mm}
    \label{tab:overlap}
\end{table*}

\begin{table}[t]
\scalebox{0.75}{
    \centering
    \begin{tabular}{c|ccccc}
    \hline
       & AG news & {\makecell{AG news \\(full)}} & Blog & TP-US & Yelp\\
         \hline
     Victim model & 79.99 & 94.47 & 97.07 &  85.53 & 95.57\\
         \hline\hline
  \makecell{$\mathcal{T}_A=\mathcal{T}_V $\\($D_V$)} &  \textbf{80.83} & \textbf{94.54}  & \textbf{96.77} &  86.48 & \textbf{95.72} \\
        \makecell{$\mathcal{T}_A=\mathcal{T}_V$ \\($D_A$)} &  80.10 & 94.38 & 95.64 & \textbf{86.53} & 95.61\\
        \hline\hline
    \makecell{$\mathcal{T}_A \neq \mathcal{T}_V$ \\(reviews)}\\
         \hline
         0.1x & 50.90 & 86.57 &  36.83& 79.95  & 92.39 \\
        0.5x & 68.13 & 87.31 &84.59  & 84.21  & 93.25\\
       1x  &  69.94 & 88.63 & 88.16 &  85.15 & 94.06 \\
        5x &  75.29& 91.27 & 92.75 &  85.82 & 94.95 \\
        \hline\hline
         \makecell{$\mathcal{T}_A \neq \mathcal{T}_V$ \\(news)}\\
         \hline
          0.1x & 61.70 & 89.13 &18.04  & 79.20  & 88.24\\
        0.5x & 70.56 & 89.84 & 32.92 & 84.18  & 89.76\\
        1x  &  71.95  & 90.47 & 83.13	&  84.15	& 91.06 \\
        5x  &   75.82 	&92.26 & 87.64	& 85.46	& 93.13 \\
        \hline
    \end{tabular}}
    \vspace{3pt}
    \caption{Accuracy of victim models and extracted models %among different datasets in terms of 
    across different domains and query sizes. %Note the high accuracies for some tasks even at low query budgets, and diminishing accuracy gains at higher budgets.
    }
    \label{tab:mea_full}
\end{table}

\subsection{MEA Setup and Results}
\label{sec:mea_setup}

\textbf{Attack Strategies}.
Similar as~\citet{krishna2019thieves}, in this paper, we mainly study model extraction through simulated experiments. We assume that the attacker has access to the freely available pretrained BERT model used by the victim model.

\textbf{Query Distribution}.
To investigate the correlation between the query distribution ($\mathcal{T}_A$) and the effectiveness of our attacks on the victim model trained on data from distribution $\mathcal{T}_V$ (\cf\tabref{tab:data}), we explore the following two different scenarios: (1) same-domain: we sample queries from task-specific original data distribution of the victim model ($\mathcal{T}_A=\mathcal{T}_V$); (2) cross-domain: we sample queries from different distribution but semantically-coherent data as the original data ($\mathcal{T}_A \neq \mathcal{T}_V$).

For the first scenario, we use $D_Q$ as the queries.
Since the owners of APIs tend to use the in-house datasets, it is difficult for the attacker to know the target data distribution as a prior knowledge. Therefore, the second scenario is more realistic.
Specifically, we investigate the query distribution $\mathcal{T}_A$ sourced from the following corpora:

\begin{compactitem}
    \item Reviews data: Yelp and Amazon reviews dataset~\cite{zhang2015character}. {\color {black}It is worth noting that we exclude Yelp reviews dataset from the Yelp task to guarantee a fair evaluation.} 
    \item News data: CNN/DailyMail dataset~\cite{hermann2015teaching}.
\end{compactitem}

Regarding the experiments of MEA, our general findings from \tabref{tab:mea_full} include: (1) using queries from the same data distribution achieves the best extraction performance, validating that the success of the extraction is positively proportional to the domain closeness between the victim training data and queries; (2) using queries from the same data distribution, the extracted model can achieve comparable accuracies, even outperform the victim models. We hypothesize this is due to the regularizing effect of training on soft-labels~\cite{hinton2015distilling}; (3) our MEA is still effective despite the fact that queries may come from different distributions. Using samples from different corpora (review and news) as queries, our MEA can still achieve 0.85-0.99$\times$ victim models' accuracies when the number of queries varies in \{1x,5x\}.

We also noticed that AG news prefers \textit{news} data, while \resource{tp-us}, Blog and Yelp prefer \textit{reviews} data. Intuitively, one can attribute this preference to the genre similarity, \ie \textit{news} data is close to AG news, while distant from \resource{tp-us}, Blog and Yelp. To rigorously study this phenomenon, we calculate the uni-gram and 5-gram overlapping between test sets and different queries in the 1x setting, and corroborate that there is a positive correlation between the accuracy and the lexicon similarity. \tabref{tab:overlap} corroborates that there is a positive correlation between the accuracy and the lexicon similarity. 
From now, unless otherwise mentioned, we will use \textit{news} data as queries for \resource{ag} news, and \textit{reviews} data as queries for \resource{tp-us}, Blog and Yelp, to simulate a more practical scenario.\footnote{Empirically, we do not have access to the training data of the victim model.}

\textbf{Query Size.} For BERT-based API as a service, since each query incurs a certain expense, due to the budget limit, attackers cannot issue massive requests. We vary query size from \{0.1x,0.5x,1x,5x\} size of the training data of the victim model respectively. According to \tabref{tab:mea_full}, although more queries indicate a better extraction performance, small query budgets (0.1x and 0.5x) are often sufficiently successful. Despite some datasets such as Blog may suffer from a drastic drop, the overall performance of the extracted models are comparable to the victim models. 

\begin{table}
\scalebox{0.85}{
% \small
% \begin{center}
% \begin{wraptable}[]
     \centering
    \begin{tabular}{cccc}
         \hline
       & AG news & Blog & TP-US\\
         \toprule
Majority class & 49.94 & 49.57 & 38.15 \\
\midrule
BERT (w/o fine-tuning)&  69.39 &  44.03 & 49.38\\
\midrule
$\mathcal{T}_A=\mathcal{T}_V$ ($D_V$) &   22.74  &  36.37&  36.76\\
\midrule
$\mathcal{T}_A = \mathcal{T}_V$ ($D_A$)\\

1x  &   21.01  & 35.98 &  37.34\\
\midrule
$\mathcal{T}_A \neq \mathcal{T}_V$ (reviews)\\
       1x  &   17.93 &  34.34  & \textbf{35.97} \\
        5x &  18.31 &  34.45&  36.82 \\
\midrule
$\mathcal{T}_A \neq \mathcal{T}_V$ (news)\\
        1x  &    \textbf{15.76} & \textbf{33.88}	& 36.92 \\
        5x  &   17.91	&  35.39	&37.68 \\
% \midrule
% $D_A \rightarrow D_V$ &   18.61 &   34.78  &37.70 \\
\bottomrule
    \end{tabular}}
    \vspace{3pt}
    \caption{%Privacy protection of attribute inference %(privacy) 
    AIA attack success over different datasets. The extracted model is trained on the queries from different distributions, and the attribute inference model is trained on the public data $D_A$ (\cf Algorithm~\ref{Algorithm:aia}). Note higher value means better empirical privacy.%, \ie lower attack success. %(2nd half: $D_A$ reserved for the training of AIA attack model.% to mount AIA).
    }
    % \vspace{-20pt}
    \label{tab:aia}
\end{table}
%\vspace{-5mm}
% \end{wraptable}

\begin{table}[t]
\scalebox{0.535}{
    \centering
    \begin{tabular}{cccccc}
    \hline
     \multicolumn{6}{c}{\resource{ag} news}\\
     \hline
        	& {\makecell{entity 0 \\(std=310.0)}}	& {\makecell{entity 1 \\(std=1568.5)}}	&{\makecell{entity 2 \\(std=2095.5)}}	& {\makecell{entity 3 \\(std=2640.5)}}	& {\makecell{entity 4 \\(std=2615.5)}}\\
        	\hline
$\mathcal{T}_A=\mathcal{T}_V$	& 15.61 &    15.10	&  7.71	&   6.95	& 5.49 \\
\makecell{$\mathcal{T}_A \neq \mathcal{T}_V$ \\(news)} 
& 14.79	& 12.38	& 3.84	& 5.33	&2.02\\ \hline
% \multicolumn{5}{c}{} \\
% \hline
& \multicolumn{2}{c|}{\resource{tp-us}} & \multicolumn{2}{c}{Blog} \\
\hline
	& {\makecell{gender \\(std=1512.0)}}	&\multicolumn{1}{c|}{{\makecell{age \\(std=1440.0)}}} %&		
	&{\makecell{age \\(std=28.0)}}	&{\makecell{gender \\(std=6.0)}} \\
$\mathcal{T}_A=\mathcal{T}_V$	& 36.40& \multicolumn{1}{c|}{37.12}	& %$\mathcal{T}_A=\mathcal{T}_V$	& 
32.18	& 39.02\\
\makecell{$\mathcal{T}_A \neq \mathcal{T}_V$ \\(reviews)}&	36.44	&\multicolumn{1}{c|}{37.40}	& %$\mathcal{T}_A \neq \mathcal{T}_V$ \\(reviews) &
31.20	&38.01\\
\hline
    \end{tabular}}
    \vspace{3pt}
    \caption{%The accuracy of the individual attributes of the attackers on extracted models 
    AIA performance on attributes of different datasets. All experiments are based on 1x queries. std is the standard deviation of attribute distribution.}
    \label{tab:att_acc}
\end{table}

%\begin{wraptable}{r}{7cm}

\begin{table}
\begin{center}
\scalebox{0.75}{
\begin{tabular}{lllll} 
 \toprule
 \textbf{Victim} & \textbf{Extracted} & \multicolumn{3}{c}{\textbf{TP-US}}\\
 \cmidrule(lr){3-5}
 & & victim $\uparrow$ &MEA $\uparrow$ & AIA $\downarrow$ \\
 \midrule
 BERT-large & BERT-base & 86.82 & 85.36 &36.65 \\
 RoBERTa-large & BERT-base & 87.20 & 85.72 & 37.33 \\
 RoBERTa-base & BERT-base & 86.66 & 85.40 &37.52 \\
 XLNET-large & BERT-base & 87.21 & 85.99 &37.68 \\
 XLNET-base & BERT-base & 86.91 & \textbf{86.13} &38.09 \\
 \midrule
   BERT-base & BERT-base & 85.53 & 85.15 & \textbf{35.97} \\
%  BERT-large & BERT-large & 86.82 & 86.23 & \textbf{35.08} & 94.75 & \textbf{91.01} & \textbf{59.3}\\
%  BERT-base & BERT-large & * & 85.89 & 36.89 & * & 90.68& 37.2\\
%  BERT-large & BERT-base & * &  85.36 &36.65 & * & 89.88&42.7 \\
 \bottomrule
\end{tabular}}
\end{center}
\vspace{3pt}
\caption{%Test accuracy using queries on TP-US, AG and Blog 
Attack performance on TP-US with mismatched architectures between the victim model and the extracted model. %Note the trend: (large, large) $>$ (base, large) $>$ (base, base) $>$ (large, base) where the 
%($\cdot$, $\cdot$) refers to (victim, extracted) pretraining.
}
\vspace{-10pt}
\label{tab:bert_mismatch}
\end{table}
%\end{wraptable}

\subsection{AIA Setup and Results}
\label{sec:aia}
For AIA, we only study TP-US, \resource{ag} news and Blog datasets, as there is no matching demographic information for Yelp. Following~\cite{coavoux2018privacy, lyu2020differentially}, for demographic variables (\ie gender and age), we take $1-X$ as \textit{empirical privacy}, where $X$ is the average prediction accuracy of the attack models on these two variables; for named entities, we take $1-F$ as \textit{empirical privacy}, where $F$ is the F1 score between the ground truths and the prediction by the attackers on the presence of all the named entities. Higher empirical privacy means lower attack performance.

On the extracted model from MEA, attackers can determine how to infer the private attributes from the BERT representation $\vh$ of the extracted model. For each attribute of interest, we train an attribute inference model. Each attack model consists of a multi-layer feed forward network and a binary classifier, which takes $\vh$ as the inputs and emits the predicted attribute. Once the attribute inference
model is obtained, we measure the empirical privacy by the ability of the model to predict accurately the specific private attribute in $D_V$.

To gauge the private information leakage, we consider a majority
value for each attribute as a baseline. To evaluate whether our extracted model can help enhance AIA, we also take the pretrained BERT without (w/o) fine-tuning as a baseline. Note that BERT (w/o fine-tuning) is a plain model that does not contain any information about the target model training data.

We show AIA results using both the same-domain and cross-domain auxiliary data in \tabref{tab:aia}. In particular, we set a \textbf{limited} amount of the same-domain auxiliary data as the non-overlapping 10\% original data (Section~\ref{sec:data_partition}), and sample the cross-domain auxiliary data from the cross-domain queries in Section~\ref{sec:mea_setup}. \tabref{tab:aia} shows that compared to the BERT (w/o fine-tuning) and majority baseline, the attack model built on the BERT representation of the extracted model 
indeed largely enhances the attribute inference for the victim training data — more than 3.57-4.97x effective for AG news compared with the majority and BERT (w/o fine-tuning) baselines, even when MEA is trained on the queries from different data distributions. The reason is quite straightforward: the extracted model is trained on the queries and the returned predictions from the victim model, while BERT (w/o fine-tuning) is a plain model that did not contain any information about the victim training data, and the majority baseline is merely a random guess. Hence, it is not surprising that AIA based on the pretrained BERT can consistently outperform both baselines. 
This also implies that the target model predictions inadvertently capture sensitive information about users, such as their gender, age, and other important attributes, apart from the useful information for the main task (\cf~\tabref{tab:mea_full}).

%=====================================================
\begin{figure*}[tb]
	\centering
	\subfloat[\resource{ag} news]{\includegraphics[width=1.8in]{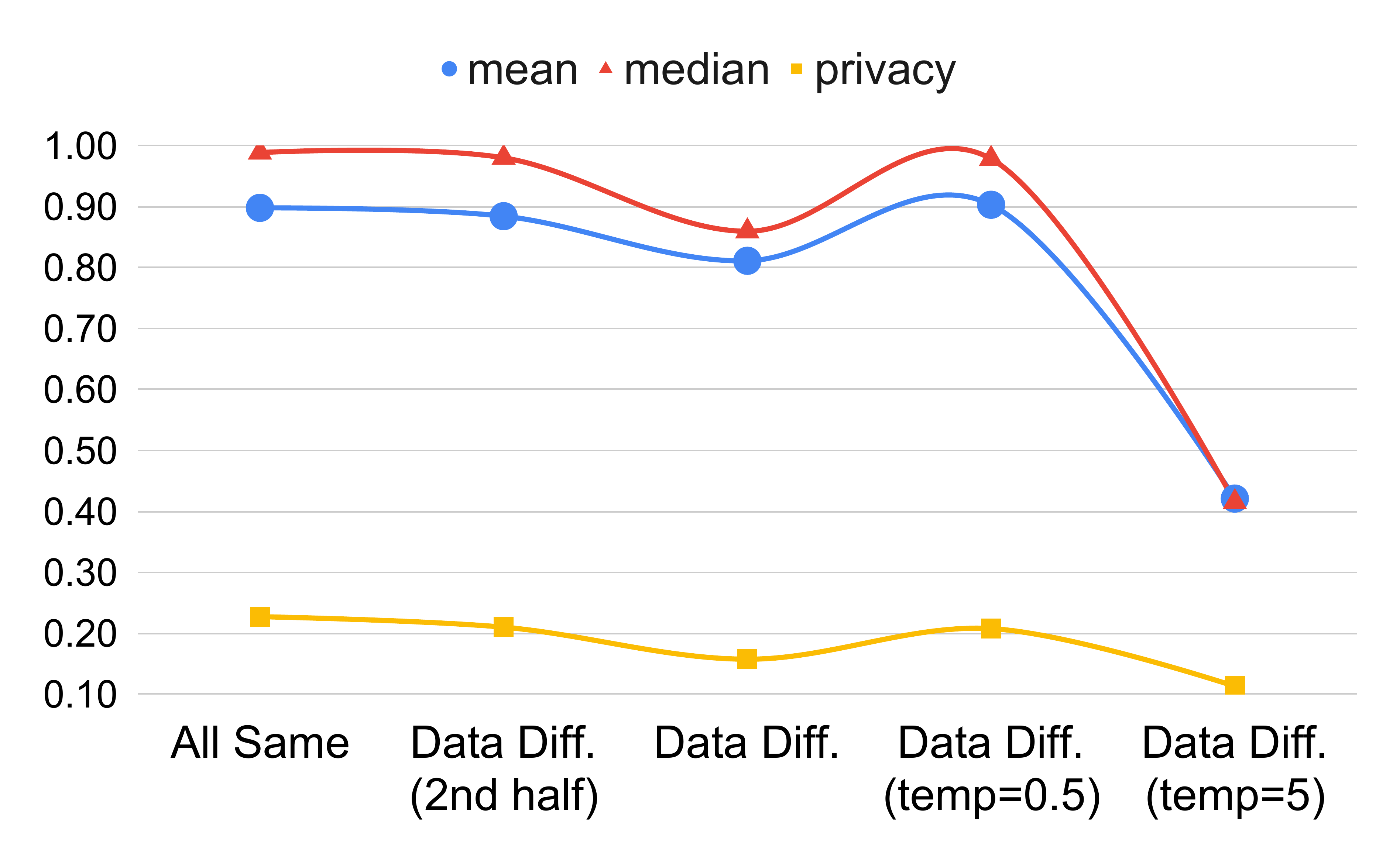}}\ \ \
	\subfloat[\resource{tp-us}]{\includegraphics[width=1.8in]{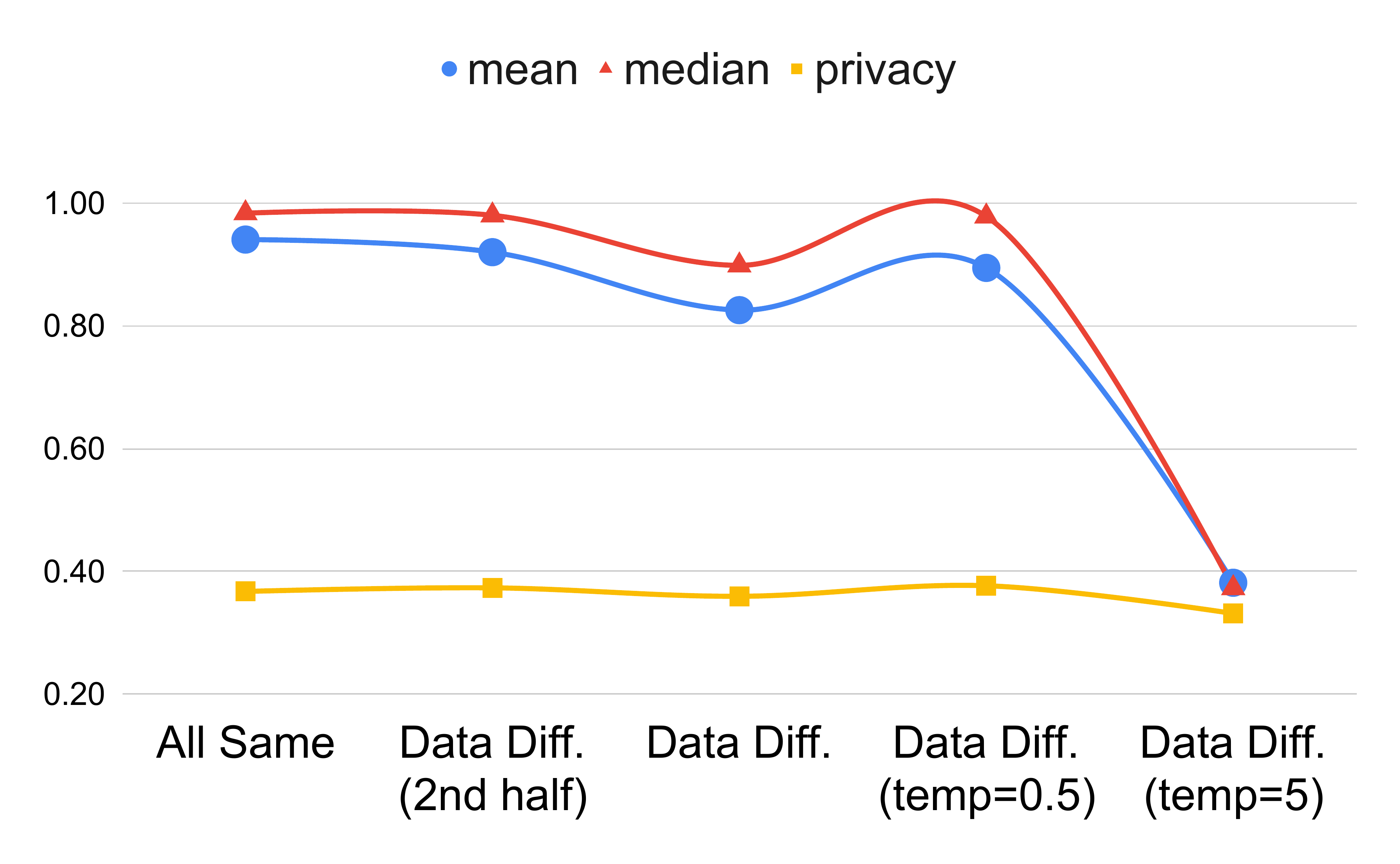}}\ \ \
	\subfloat[Blog]{\includegraphics[width=1.8in]{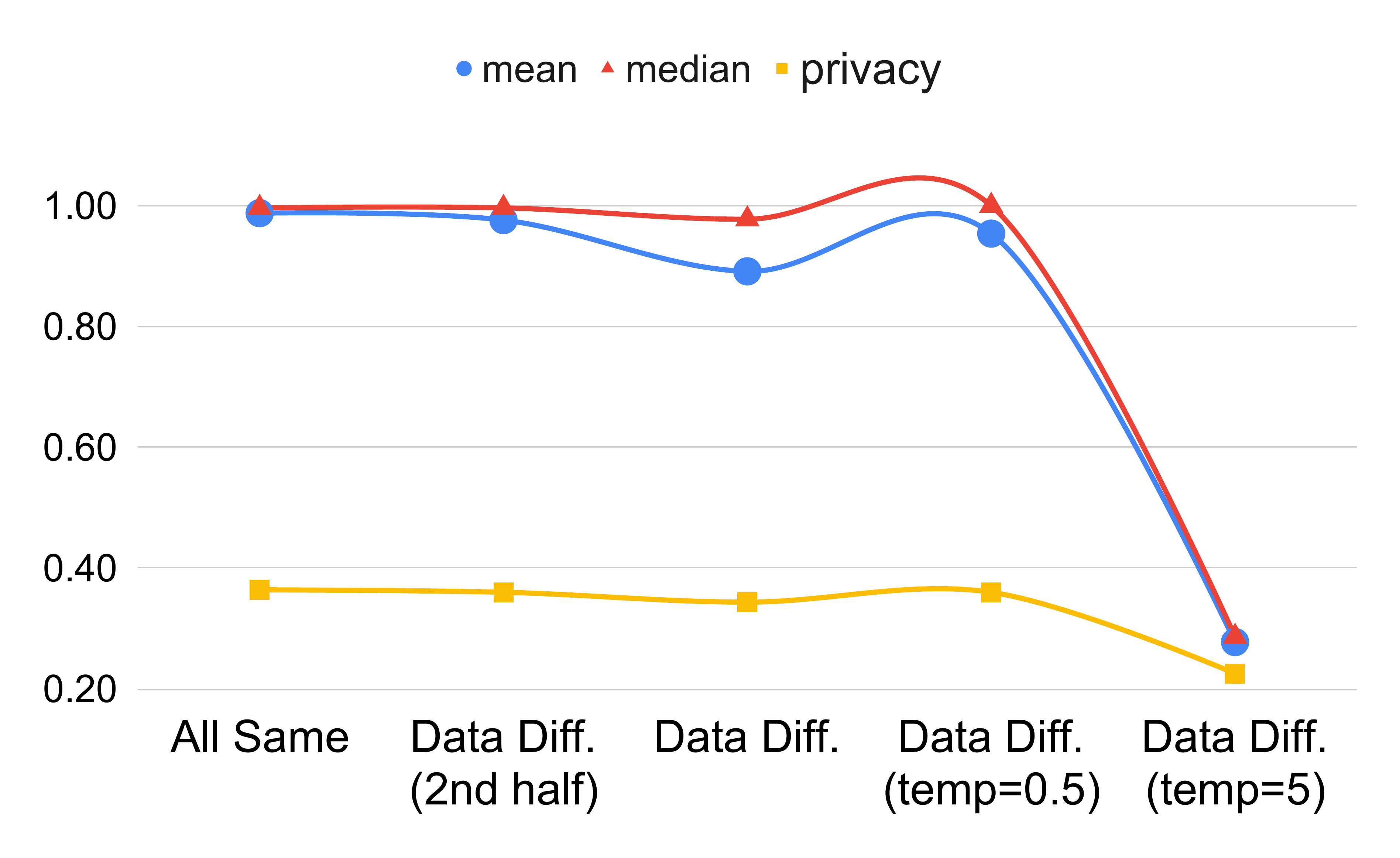}}\\
	\caption{The correlation between the empirical privacy of AIA and the maximum posterior probability. \resource{ag} news uses news data as queries, while Blog and \resource{tp-us} query the victim models with reviews data. \textbf{mean} and \textbf{median} denote the mean and median of the maximum posterior probability of the queries.}
    \label{fig:aia_maxprob}
\end{figure*}
%======================================================

\textbf{Impact of prediction sharpness}. Interestingly, compared with the queries from the same distribution, \tabref{tab:aia} also shows that queries from different distributions make AIA easier (see the best results corresponding to the lower privacy protections in bold in \tabref{tab:aia}).
We hypothesize this counter-intuitive phenomenon is due to that the posterior probability of the same distribution is sharper than that of the different distribution.
This argument can be further strengthened in Section \ref{sec:defense}, in which we use a temperature coefficient $\tau$ at the softmax layer to control the sharpness of the posterior probability. We conjecture that if the model is less confident on its most likely prediction, then AIA is more likely to be successful. This speculation is confirmed by \figref{fig:aia_maxprob}, where the higher posterior probability leads to a higher empirical privacy.

\textbf{Impact of attribute distribution}. We further investigate which attribute is more vulnerable, \ie the relationship between attribute distribution (histogram variance) and privacy leakage. 
\tabref{tab:att_acc} empirically indicates that attributes with higher variances cause more information leakage or a lower empirical privacy. For example, for AG-news, entity 2-4 with higher variances result in lower empirical privacy, while entity 0-1 are more resistant to AIA. For TP-US and Blog, as age and gender exhibit similar distribution, AIA performance gap across these two attributes is less obvious.

\begin{table*}[t]
\scalebox{0.9}{
    \centering
    % \resizebox{\columnwidth}{!}{%
    \begin{tabular}{llccccccccc}
\hline
\multirow{2}{*}{} & &\multicolumn{3}{c}{\resource{ag} news} & \multicolumn{3}{c}{\resource{blog}}& \multicolumn{3}{c}{\resource{tp-us}}\\
\cmidrule(lr){3-5}  \cmidrule(lr){6-8}  \cmidrule(lr){9-11} 
       & & Utility$\uparrow$ & MEA$\uparrow$ &AIA $\downarrow$ & Utility$\uparrow$ & MEA  $\uparrow$ &AIA  $\downarrow$  & Utility$\uparrow$ & MEA  $\uparrow$ &AIA  $\downarrow$\\
\toprule
%\resource{majority} & 79.40  & 36.39 & 57.79  & 49.34 & 34.16  & 46.96\\
%\midrule
\multicolumn{2}{c}{\resource{ND}} &79.99 & 71.95 & 15.76 &\textbf{97.07} & 88.16& 34.34 & 85.53& 85.33 & 36.92 \\
\midrule
%\resource{PP}& *  & * & *  & * &  * & - & * &  * & *\\
% \resource{soft.}\\
\multirow{2}{*}{{\rotatebox{90}{\resource{soft.}}}}
& \quad $\tau=0.0$ & 79.99 & 69.11 & \textbf{22.47} &97.07 & 85.57 & \textbf{35.19} & 85.53& 84.60 & 37.62  \\
& \quad $\tau=0.5$ & 79.99 & 72.32 & 20.78 &97.07 & 85.68 &  34.91& 85.53& 85.10 & \textbf{37.69} \\
& \quad $\tau=5$ &79.99 & 72.48 & 11.32 & 97.07 & 86.73 & 33.80&85.53 & 85.33 & 33.18 \\
\midrule
% \resource{pert.} & 	& 	&& \\
% \resource{top-5} & -0.25 & +0.67 & -21.71 &+26.43 &-2.44  &+1.16\\
% \resource{dpnr} & \bfseries +0.12 &\textbf{+3.66} & \textbf{+2.12} & \textbf{+31.13} &-0.38 &\textbf{+15.86}\\
\multirow{2}{*}{{\rotatebox{90}{\resource{pert.}}}}
& \quad $\sigma=0.05$ &\textbf{80.03} & 71.47 & 14.46  & 96.17 & 85.87 & 34.75 & \textbf{85.83} & 85.09 & 37.43\\

& \quad $\sigma=0.2$  & 79.41 & 71.61 & 12.58 &  95.38 & 85.31 & 34.97 & 85.65 & 84.98 & 36.90\\
& \quad $\sigma=0.5$ &65.13 & \textbf{69.05} & 11.66 & 62.23 & \textbf{81.77} & 33.79 & 63.21 & \textbf{83.88} & 35.90\\
\bottomrule
\end{tabular}%
}
\vspace{3pt}
\caption{Attack performance under different defenses and datasets. ND: no defense; $\tau$: temperature on softmax. $\sigma$: variance of Gaussian noise. Utility means the accuracy of the victim model after adopting defense. For MEA, \textbf{lower} scores indicate better defenses, %while it is \textbf{inverse} 
conversely for AIA. %We use white-box attack for AET. 
All experiments are conducted with 1x queries.}
%PP: prediction poisoning; }
    \label{tab:defense}
%    }
\end{table*}

\subsection{Architecture Mismatch}
\label{sec:arch_diff}

{\color{black}In practice, it is more likely that the adversary does not know the victim's model architecture. A natural question is whether our attacks are still possible when the extracted models and the victim models have different architectures, \ie architectural mismatch. To study the influence of the architectural mismatch, we fix the architecture of the extracted model, while varying the victim model from BERT~\cite{devlin2018bert}, RoBERTa~\cite{liu2019roberta} to XLNET~\cite{yang2019xlnet}. According to \tabref{tab:bert_mismatch}, when there is an architecture mismatch between the victim model and the extracted model, the efficacy of AIA is alleviated as expected. However, the leakage of the private information is still severe (\cf the majority class in \tabref{tab:aia}). Surprisingly, we observe that MEA cannot benefit from a more accurate victim, which is different from the findings in~\cite{hinton2015distilling}. For example, the victim model performs best using XLNET-large, while MEA shows best attack performance when the victim model uses XLNET-base. We conjecture such difference is ascribed to the distribution mismatch between the training data of the victim model and the queries. We will conduct an in-depth study on this in the future.}

\section{Defense}
\label{sec:defense}
Although we primarily focus on the vulnerabilities of BERT-based APIs in this work, we briefly discuss two potential counter strategies the victim model may adopt to reduce the informativeness of prediction while maintaining API performance.

\textbf{Softening predictions (SOFT).} For soft label, a temperature coefficient $\tau$ on softmax layer manipulates the distribution of the posterior probability. A higher $\tau$ leads to a smoother probability vector, whereas a lower one produces a sharper distribution. When $\tau$ is approaching 0, the posterior probability becomes a hard label.

\textbf{Prediction perturbation (PERT)}. Although the temperature term used for softmax can transform the posterior probability, it does not affect the predicted label, \ie $\mathrm{argmax}(\vy)$ is not changed.
One can leverage an additive noise $\vn$ to pollute the predictions~\cite{lyu2020differentially_SIGIR}. Such noise can be sampled from Gaussian distribution with a variance of $\sigma$. Thus, a perturbed posterior probability with sum=1 can be obtained via $\vy' = \textrm{normalize}(\vy + \vn)$.

\tabref{tab:defense} indicates that varying temperature on softmax cannot well defend against MEA. In some cases, softening predictions can even help improve the attack performance. For example, when applying $\tau=5$ on the prediction vector of AG news, the extracted model outperforms the extracted model trained on the original prediction vector. Meanwhile, compared with no defense (ND), hard label can slightly mitigate the information leakage caused by AIA.
Overall, models can still be effectively stolen and exploited using just the hard label or the smoothed predictions returned by the black-box API. This further corroborates that the adversary does not always need the confidence scores for our attacks to be successful. 

On the other hand, according to \tabref{tab:defense}, prediction perturbation can effectively protect APIs from MEA, only when the performance of APIs severely degrades. Moreover, such degradation is generally larger than that to attackers. Consequently, the extracted model may surpass the victim model. On the contrary, AIA is a beneficiary of the perturbation. For example, a larger perturbation imposes more privacy leakage on the victim model.

\section{Conclusions}
This work goes beyond model extraction from BERT-based APIs, we also identified that the extracted model can largely enhance the privacy leakage even in difficult scenarios (\eg limited query budget; queries from distributions that are different from the training data distribution of the victim APIs; when the architecture of the victim model is unknown, etc). Extensive experiments based on the representative NLP datasets and tasks under various settings demonstrate the effectiveness of our attacks. In future, we expect to extend our work to more complex NLP tasks, and develop more effective defenses. 

\bibliographystyle{acl_natbib}
\bibliography{sample-base}

\end{sloppypar} 
\end{document}